\documentstyle[epsfig,12pt]{article}

\newlength{\dinwidth}
\newlength{\dinmargin}
\begin{document}

% \draft
\title{Jamming Non-Local Quantum Correlations}
\author{Jacob Grunhaus, Sandu Popescu and Daniel Rohrlich\\[1cm]
School of Physics and Astronomy\\
Tel-Aviv University\\
Ramat-Aviv 69978 Tel-Aviv, Israel}
\date{}
\maketitle

\vspace{-12.5cm}
\bigskip
\begin{tabbing}
\` {\large TAUP-2263-95} \\
\` June 21, 1995 \\
\end{tabbing}
\bigskip
\vspace{8.2cm}

\begin{abstract}
We present a possible scheme to  tamper with non-local
quantum correlations in a way that is consistent with relativistic
causality, but goes beyond quantum mechanics.
A non-local ``jamming" mechanism, operating within a
certain space-time window, would not violate relativistic
causality and would not lead to  contradictory causal loops.
The results presented in this Letter do not depend on any model of how
quantum correlations arise and apply to any jamming mechanism.
\end{abstract}
\section{Introduction}
The question of non-local quantum correlations versus local realism,
first raised in the famous EPR paper \cite{epr}, has held the interest
of the physics community since.  J. S. Bell \cite{bell} showed that
the predictions of quantum mechanics are incompatible with any model
based on local realism.  The pioneering experimental work of A. Aspect
{\it et al.} \cite{aspect} and others \cite{mandel}
supports the
predictions of quantum mechanics and contradicts local realism:
Bell inequalities applicable to the various experimental arrangements
have been shown to be violated.  It should be mentioned that
 some aspects of the
experimental setups have been criticized and questioned
\cite{santos}. Problems of  experimental
bias or enhancement of particular polarization states by detection
systems were experimentally checked by T. Haji-Hassan {\it et al.}
\cite{hassan}
and found absent. And more recently Kwiat {\it et al.} \cite{kwiat} have
proposed and described an experimental arrangement that overcomes
shortcomings of previous experiments.  While experiments are still open
to criticism, it is generally accepted that local realism
is untenable. In this Letter we
  assume that in nature there exist non-local
   correlations, as predicted by quantum mechanics,
 and we address the following
 question:  Can an experimenter {\it non-locally tamper}
with non-local correlations, without violating relativistic causality?

     Quantum mechanics predicts non-local correlations; however, it does
not provide an ``explanation"  about what creates them.
  Several theoretical
models go beyond quantum mechanics and propose to explain
the phenomenon of non-local correlations via a superluminal
``communication link'' \cite{bohm}.  If one accepts the possibility of
a communication link, then a natural next step would be to
probe whether it is possible to tamper  with this link and {\it jam}
the superluminal communication \cite{shimony}.

Up to now, the possibility of jamming non-local
correlations has not received due consideration, perhaps because of
a tacit assumption that such tampering necessarily violates
relativistic causality.
(The expression {\it relativistic causality} is used here to
denote the principle that information cannot be transferred at speeds
 exceeding the speed of light.)
 In this Letter we show that jamming of
non-local correlations can be consistent with relativistic
causality.  Our results are independent of the model used to describe
how the non-local quantum correlations arise, that is, the
 nature of the
superluminal communication link, and they apply to any jamming
 mechanism.
\section{The Jamming Scheme}
Jamming might take many forms. The following  discussion does
  not discuss a mechanism for jamming; rather, it defines the
 constraints that any jamming mechanism
   must obey in order to be consistent with relativistic causality.
 In order to derive and
illustrate the constraints, it is convenient to consider a particular
experimental arrangement which can be subjected to jamming
\cite{generic}.
 We will
consider an EPR-Bohm experimental arrangement
  to study pairs of spin-$1/2$
particles entangled in a singlet state \cite{eprb}.  Spacelike separated
spin measurements on these pairs allow a test of the Bell inequalities.
Suppose that two experimenters, Alice and Bob, perform
the spin measurements.  One particle of each entangled pair arrives at
Alice's analyzing station and the other particle arrives at Bob's.
When Alice and Bob get together and combine the  results of their
measurements, they will find  violations of the Bell inequalities,
as predicted by quantum mechanics \cite{bell}.

 We now introduce a third experimenter, Jim, the jammer, who has
access to a jamming device which  he can activate, at will, and tamper
with
the communication link between each entangled pair of particles.
His action is spacelike separated from the measurements of
 Alice or  Bob or from both of them.
Jamming acts at a distance to modify the correlations between
the particles; it disturbs
 the conditions  which make possible the phenomenon of
non-local quantum correlations.  Therefore, the correlations
measured jointly by Alice and Bob will not agree with the predictions
of quantum mechanics.

     Jamming is truly non-local and cannot be carried out within the
framework of
quantum mechanics. For example,
 consider three systems, $S_1$, $S_2$ and $S_3$,
in a quantum state $\Psi_{123}$.  Let experimenters near
$S_1$ and $S_2$ measure $A^{(1)}$ and $A^{(2)}$, with eigenstates
 denoted by
$\vert a^{(1)}_i \rangle$ and $\vert a^{(2)}_j \rangle$, respectively.
The only freedom  available to an experimenter near
 $S_3$ is the choice of what local operator
$A^{(3)}$ to measure.  But the probabilities $P(a^{(1)}_i ,a^{(2)}_j )$
 for
outcomes $A^{(1)} =a^{(1)}_i$ and $A^{(2)} =a^{(2)}_j$,
\begin{equation}
P(a^{(1)}_i ,a^{(2)}_j )
=\sum_k \vert \langle \Psi_{123} \vert a^{(1)}_i
,a^{(2)}_j ,a^{(3)}_k \rangle
\vert^2~,
\end{equation}
are {\it independent} of the choice of operator $A^{(3)}$. Thus
no measurement on $S_3$ can affect the results of the measurements
performed on  $S_1$
and $S_2$, even if the three systems have interacted in the past
\cite{note}.

In general, jamming would allow Jim to send superluminal signals.
The constraints that
 must be  satisfied  in order to insure that
 Jim cannot send superluminal signals are embodied in two conditions.
The first condition,
the {\it unary condition}, a necessary but
 not sufficient  condition, requires that Jim not be able to send
  signals
 to Alice or  Bob {\it separately}. In effect
  this condition demands that
 Alice and Bob, separately, measure
 zero average spin along any axis.  Explicitly, let $N_a (+)$ and
$N_a (-)$ tally the number of spin-up and spin-down results,
 respectively,
found by Alice for a given axis.  For the same axis, let $n(k,l)$ tally,
 in the absence of jamming,
the joint results of Alice and Bob. The parameters
 $k$ and $l$ denote, respectively, the results
 ( $+$ or $-$ ) of the
  polarization measurements
  carried out by Alice and Bob.
Let $n^\prime (k,l)$ tally, in the presence of jamming, the
 corresponding polarization measurements carried out by Alice and Bob.
 The unary condition imposes the
following relations between $n(k,l)$ and $n^\prime (k,l)$:
\begin{eqnarray}
\label{Na}
N_a (+) &=& n(+,+) + n(+,-) = n^\prime (+,+) + n^\prime (+,-)\nonumber \\
N_a (-) &=& n(-,+) + n(-,-) = n^\prime (-,+) + n^\prime (-,-).
\end{eqnarray}
A similar set of relations  holds for the results $N_b (+)$ and
$N_b (-)$ found by Bob.  Hence regardless of
  whether  Jim has activated the
jamming device, Alice and Bob will find that the average spin projection
along any axis tends to zero, and Jim cannot send superluminal signals,
separately, to either Alice or Bob.

 The unary condition
allows a range of possibilities for the
 jammed  correlations: from
correlations which are only slightly different from those predicted
by   quantum mechanics, down to completely random
correlations. In particular, the unary condition allows
conservation of angular momentum, {\it i.e.} perfect anticorrelation
of spin components along any parallel axes.
\section{The Space-Time Window}
   As stated in the previous section,
 the unary condition is a necessary but not sufficient condition.  For
jamming to respect relativistic causality, we must also restrict the
relationships in space and time among the three events $a$, $b$ and $j$
generated, respectively, by Alice, Bob and Jim.  Fig. 1
 shows the geometry of
three different configurations of an EPR-Bohm experimental setup along
with  the corresponding Minkowski diagrams of the events
$a$, $b$ and $j$. In the configuration shown in
 Fig. 1(a), jamming is
{\it not} permitted.  Here Alice and Bob are in close proximity while
Jim is far away.  If jamming were permitted, Alice and Bob could
---immediately after Jim activates the jamming device---measure the spin
projections of their respective particles and combine their results
to determine the spin correlations.  They would find spin correlations
differing from the predictions of quantum mechanics  and infer that
Jim activated the jamming device. The corresponding Minkowski
diagram, Fig. 1(b), shows that
 the future light cones of $a$ and $b$ overlap,
in part, outside the future light cone of $j$.  A light signal
originating at $j$ cannot reach this overlap region
 of $a$ and $b$,
where Alice and Bob can combine their results. Were jamming
possible here, it would violate relativistic causality.

Fig. 1(c) shows a configuration that would  also permit
superluminal signalling:  Jim obtains the results of Alice's
  measurements prior to
deciding whether to activate the jamming device.  Bob is far from both
Alice and Jim.  The corresponding Minkowski diagram, Fig. 1(d), shows
that $a$ precedes $j$ by a timelike interval and both $a$ and $j$ are
spacelike separated from $b$.  Since Jim has access to Alice's results,
he can send a superluminal signal to Bob by {\it selectively} jamming:
For instance, suppose Jim activates the jamming device only when Alice
obtains the value $ + 1/2$ for the projection of the
  spin of a particle.   Bob will, then, find that
the average spin component along a given axis does {\it not} tend to
zero. The preceding can be demonstrated
 by comparing the results of the spin measurements
$N_b (+)$ and $ N_b (-)$, carried out by Bob in the absence of jamming,
Eqs. (3), and in the presence of selective jamming, Eqs. (4).
The notation
previously defined is used in Eqs. (3-4).
\begin{eqnarray}
\label{Nb}
N_b (+) &=& n(+,+) + n(-,+) \nonumber \\
N_b (-) &=& n(+,-) + n(-,-) ~,
\end{eqnarray}

\begin{eqnarray}
\label{sel}
N_b (+) &=& n^\prime (+,+) + n(-,+) \nonumber \\
N_b (-) &=& n^\prime (+,-) + n(-,-)~.
\end{eqnarray}

Hence the results obtained by Bob in the presence of selective
jamming will be different from those obtained in the
  absence of jamming unless
$n^\prime(+,+) =n(+,+)$ and $n^\prime(+,-) =n(+,-)$. However, the
latter requirements imply
  that jamming, in this configuration, can not have any
   discernible effect, i.e. jamming in this configuration is
   impossible.

     To eliminate configurations which
  allow violations of
  relativistic causality, as shown in Fig. 1(a) to Fig. 1(d),
we further restrict jamming by
 imposing a second condition, the {\it binary
condition}. The binary condition, which is manifestly covariant,
demands that  the overlap of the future light cones
of $a$ and $b$  lie entirely within the future light cone
of $j$ and therefore a light signal emanating from $j$ can
 reach  the overlap region.
The configuration shown in Fig. 1(a) and 1(b), which allows an overlap of
the future light cones of $a$ and $b$ outside of the future light
cone of $j$, is therefore forbidden. The configuration shown in
Fig 1(c) and 1(d), a configuration for
 selective jamming,  violates the unary condition
and it is also disallowed by the

\bigskip

\begin{figure}
\centerline{\epsfig{file=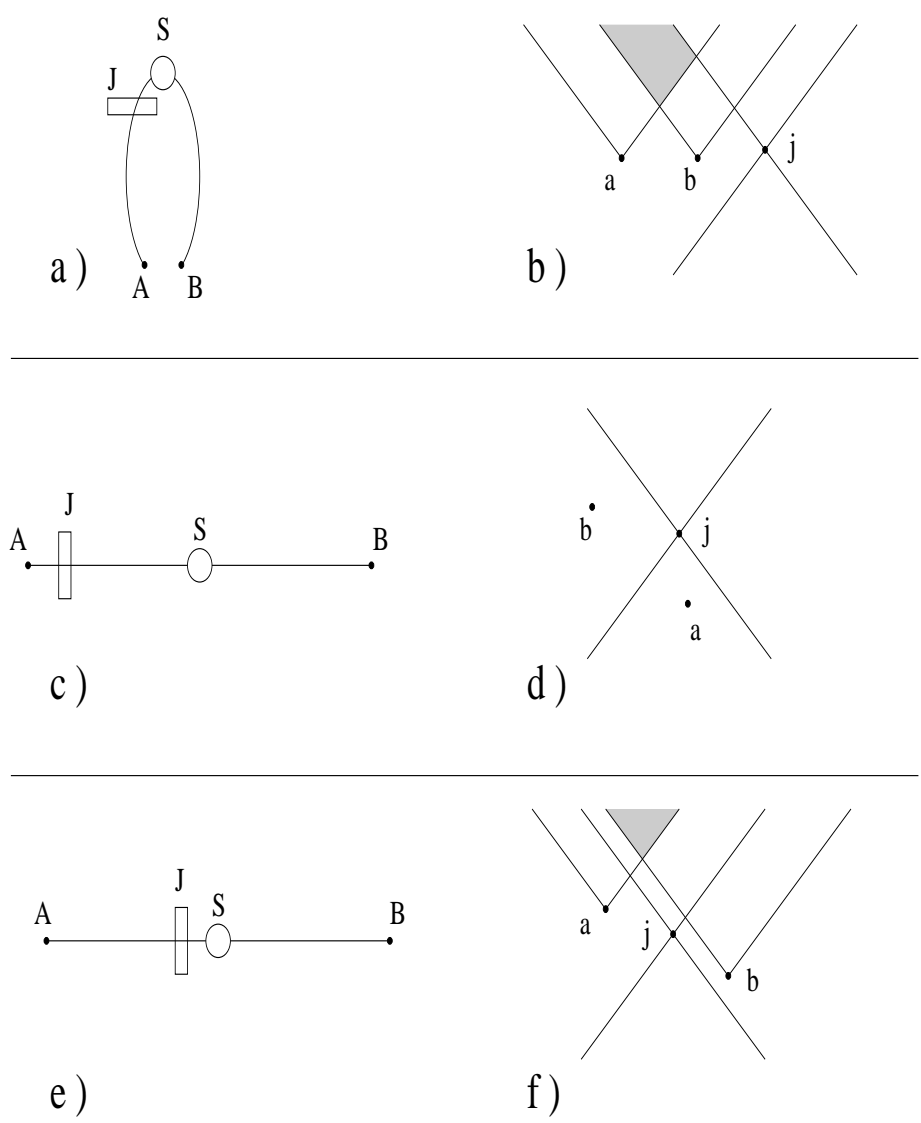}}
%\centerline{\epsfxsize = 10cm \epsfysize =15cm \epsfbox{fig88.eps}}
%\centerline{\epsffile{fig88.eps}}
\label{fig:hyp}
\caption{The geometrical configurations showing the
source $S$ of pairs of quantum systems, the jammer $J$, and the
experimenters Alice $A$, and Bob $B$.
(a) A and B are close to each other while J is far
from both of them.
(c)  A and J are close to each other while B is far
from both of them.
(e) A, B and J are all far from each other; J is stationed near the
source  and A and B are at opposite ends of an EPR-Bohm setup.
Corresponding Minkowski diagrams showing the events
$a$, $b$ and $j$. (b) The future light cones of $a$ and $b$ have some
 overlap outside the future light cone of $j$.
(d) A possible configuration for selective jamming.
(f)  A configuration satisfying the binary condition. The
future light cones of $a$ and $b$ overlap only within
the future light cone of $j$.}
\end{figure}
binary condition. A configuration which satisfies the  binary condition
is shown  in Fig. 1(e) and 1(f).
 The constraints to which a jamming configuration must conform,
in order not to violate relativistic causality,
are embodied in the unary and binary conditions.
These conditions are manifestly Lorentz
invariant. However, the time sequence of the events $a$, $b$ and $j$ is
not.  A time sequence $a$, $j$ and $b$  in one Lorentz frame may
transform into $b$, $j$ and  $a$ in another Lorentz frame.
 Hence while  one observer will claim that Alice completed
 her measurements before Jim activated his
jamming mechanism and  thus
Jim affected only the results of Bob's measurements, another observer
will claim that Bob carried out his measurements
  first and Jim affected only Alice's results. Similar situations are
encountered in quantum mechanics where different observers in
 different Lorentz frames will give conflicting interpretations
of the same set of events.  For
example, with respect to an entangled pair of particles in an EPR-Bohm
experiment, the question of  which observer caused the collapse of the
 entangled state has no Lorentz-invariant answer \cite{aa}.

If jamming is possible then one must  accept the  possibility of
reversal of the {\it cause-effect} sequence \cite{dbohm};
however, the allowed configuration
which satisfies  the {\it unary} and   {\it binary} conditions does not
lead to contradictory causal loops, i.e. no {\it effect} can send a
signal to its {\it cause}.
Indeed, consider one
jammer, $J$, who acts on the correlations between two spacelike separated
events,  $a$ and $b$.
We first recall that the
unary condition precludes signalling  to $a$ and $b$,  separately,
by $j$; therefore, only the combined results of the measurements
of $a$ and $b$ can reveal whether $J$ activated  a jamming mechanism.
In order to complete a contradictory casual
 loop one must gather the
results of the measurements of $a$ and $b$ into the past light cone of
$j$ and then send a signal to $j$, the {\it cause}.
But the binary condition requires that the overlap of the future
 light cones of
$a$ and $b$ be completely contained in the future light cone of
 $j$, so the
only place where information from $a$ and $b$ can be
put together by means of
ordinary signals is the future of $j$. One might suppose that
other jammers, using their non-local action, could somehow transmit the
information from $a$ and $b$ into the past light-cone of $j$.
Such a scheme would
require at least two
more jammers. Since these jammers must have access to the results of
$a$ and $b$, we  place $j_1$ and $j_2$ (generated by
$J_1$ and $J_2$) at timelight separations,
respectively, from $a$ and $b$. Events $a$ and $b$ are  spacelike
separated from each other and from $j$, so $j_1$  and $j_2$
will either be spacelike separated from $j$ or in its future
light cone.

The cases of $J_1$ and $J_2$ are similar, so we discuss only
$J_1$; however, the conclusions reached apply equally to
$J_1$ and $J_2$. The jammer,
$J_1$, can communicate the results of $a$  by jamming or not jamming
the non-local correlations between pairs of entangled particles
 measured at events  $a_1$ and $b_1$. Notice that in
 order to communicate the result of a single measurement
done at $a$, $J_1$ must jam (or not jam) an
 ensemble of EPR pairs. The result of a single  measurement
carried out at  $a$  is recovered from the correlations
determined from  the combined measurements done at $a_1$ and $b_1$.

For the jammer
$J_1$ to gather the information at $a$ into the past light cone of
$j$ requires that both
 $a_1$ and $b_1$ lie in the past light cone of $j$, i.e. $j$ lies in
the overlap of the future light cones of $a_1$ and $b_1$.
 This requirement, however,  is incompatible with the
   binary  condition when applied to the triplet of events,
 $a_1$,    $b_1$ and  $j_1$, which  requires that the overlap of
 $a_1$ and $b_1$ be contained within  the future light cone of  $j_1$.
 This, in turn, implies  that
 $j$ will lie in the future light cone of $j_1$,
contradicting the assumption that   $j_1$ is either
 spacelike separated from $j$ or in $j$'s future light cone.
Consequently, at least one event
  $a_1$ or $b_1$ must be spacelike
separated from $j$. Therefore the introduction of $J_1$ does not
help to gather the results of $a$ into the past light cone of $j$.
Then, by induction, we find that no  scheme
to close a contradictory causal loop, by introducing
  any number of jammers, can succeed.
\section{Conclusions}
In  quantum mechanics
non-local correlations are well established; however, these correlations
cannot be used to send superluminal signals. In this Letter we have
raised the question of whether a form of non-locality beyond quantum
mechanics---non-local tampering with quantum correlations---could also
respect relativistic causality.
We find that jamming configurations which obey
two conditions---the {\it unary} condition, which forbids
 superluminal signalling to either of two
experimenters, and the {\it binary} condition, which restricts the
space-time configuration of the two experimenters and the
jammer---respect relativistic causality. For these  configurations,
the cause-effect sequence might not be preserved in all
 Lorentz frames; however, they do not
lead to  contradictory causal loops.
Hence, we find that   a stronger form of non-locality
than that arising in quantum mechanics---action at a distance
rather than non-local correlations---is consistent with relativistic
causality. \cite{shimony,axiom,futur}

The results presented in this Letter are independent of the model used
to describe the nature of the non-local correlations and apply
to any jamming mechanism. Experimental studies, to date,
have not tested the possibility of jamming.
We suggest that current and projected EPR-Bohm experiments
test the possibility of jamming in configurations consistent with the
constraints derived in this Letter.
The constraints on the jamming configuration, however, because
of their generality, do not themselves suggest a preferred mechanism
for carrying out the jamming procedure.

We thank Y. Aharonov for helpful discussions.  The research of D. R.
was supported by the State of Israel, Ministry of Immigrant Absorption,
Center for Absorption in Science.

%\end{references}

\begin{thebibliography}{99}
%\begin{references}
\bibitem{epr}{A. Einstein, B. Podolsky and N. Rosen,
{\it Phys. Rev.} {\bf 47}, 777 (1935).}

\bibitem{bell}{J. S. Bell, {\it Physics} {\bf 1}, 195 (1964);
J. F. Clauser, M. A. Horne, A. Shimony and R. A. Holt,
{\it Phys. Rev. Lett.} {\bf 23}, 880 (1969).}

\bibitem{aspect}{A. Aspect, P. Grangier and G. Roger,
{\it Phys. Rev. Lett.} {\bf 47}, 460 (1981);
{\it Phys. Rev. Lett.} {\bf 49}, 91 (1982);
A. Aspect, J. Dalibard and G. Roger,
{\it Phys. Rev. Lett.} {\bf 49}, 1804 (1982).}


\bibitem{mandel}{Z. Y. Ou and L. Mandel,
 {\it Phys. Rev. Lett.} {\bf 61}, 50 (1988); Y. H. Shih and C. O. Alley,
 {\it Phys. Rev. Lett.} {\bf 61}, 2921 (1988).}


\bibitem{santos}{E. Santos, {\it Phys. Rev. Lett.} {\bf 66}, 1388
(1991); {\it Phys. Rev. Lett.} {\bf 68}, 2702 (1992); {\it Phys. Rev.}
{\bf A46}, 3646 (1992).}


\bibitem{hassan}{T. Haji-Hassan, A. J. Duncan, W. Perrie,
H. J. Beyer and H. Kleinpoppen, {\it Phys. Lett.} {\bf A123},
 110 (1987).}


\bibitem{kwiat}{P. G. Kwiat, P. H. Eberhard, A. M. Steinberg and
 R. Y. Chiao, {\it Phys. Rev.} {\bf A49}, 3209 (1994).}

\bibitem{bohm}{D. Bohm, {\it Wholeness and the Implicate Order}
(Routledge and Kegan Paul, London, 1980); D. Bohm and B. Hiley, {\it
Found. Phys.} {\bf 5}, 93 (1975);
 J.-P. Vigier, {\it Astr. Nachr.} {\bf303}, 55 (1982);
 N. Cufaro-Petroni and J.-P. Vigier, {\it Phys. Lett.}
{\bf A81}, 12 (1981);
 P. Droz-Vincent, {\it Phys. Rev.} {\bf D19}, 702 (1979);
 A. Garuccio, V. A. Rapisarda and J.-P. Vigier, {\it Lett. Nuovo
Cim.} {\bf 32}, 451 (1981).}

\bibitem{shimony}{A. Shimony, {\it Foundations  of Quantum Mechanics in
Light of the New Technology}, S. Kamefuchi {\it et al.}, eds. (Tokyo,
Japan Physical Society, 1983), p. 225; {\it Quantum Concepts of Space and
Time}, R. Penrose and C. Isham, eds. (Oxford, Claredon Press, 1986),
p. 182.}

\bibitem{generic}{Any entangled state of any number of systems
gives rises to non-local correlations violating a generalized
Bell inequality. See, S. Popescu and D. Rohrlich,
 {\it Phys. Lett.} {\bf A166}, 293 (1992).}

\bibitem{eprb}{D. Bohm, {\it Quantum Theory},
Prentice-Hall, New York (1951) 614.}

\bibitem{note}{Here we assumed that
 $S_1$, $S_2$ and $S_3$ are well localized. While other schemes are
possible within quantum mechanics, they, too, do not permit jamming}

\bibitem{aa}{Y. Aharonov and D. Albert, {\it Phys. Rev.} {\bf D24}, 359
(1981).}


\bibitem{dbohm}{See {\it e.g.} D. Bohm, {\it The Special Theory of
 Relativity},
W. A. Benjamin Inc., New York (1965) 156-158.}


\bibitem{axiom}{Y. Aharonov, unpublished lecture notes; S. Popescu
and D. Rohrlich, {\it Found. Phys.} {\bf 24}, 379 (1994).}


\bibitem{futur}{D. Rohrlich and S. Popescu, to appear in the Proceedings
of {\it 60 Years of E.P.R.} (Workshop on the Foundations of Quantum
Mechanics, in honor of Nathan Rosen), Technion, Israel, 1995;
S. Popescu and D. Rohrlich, to be published.}

\end{thebibliography}
\end{document}